\documentclass[useAMS,twocolumn]{emulateapj}
\pdfoutput=1
\usepackage[latin2]{inputenc}
\usepackage{hyperref}

\newcommand{\muK}{$\mathrm{\mu K}$}
\newcommand{\LCDM}{$\Lambda$CDM}

\def\zobov{{\scshape zobov}}
\def\eazy{{\scshape eazy}}
\def\swarp{{\scshape swarp}}
\def\scamp{{\scshape scamp}}

\begin{document}

\title{Galaxy Counts on the CMB Cold Spot}

\author{Benjamin R. Granett\altaffilmark{1,2}, Istv\'an Szapudi\altaffilmark{2} and Mark C. Neyrinck\altaffilmark{3}}
\altaffiltext{1}{Electronic address: granett@ifa.hawaii.edu}
\altaffiltext{2}{Institute for Astronomy, University of Hawaii, 2680 Woodlawn Drive, Honolulu HI 96822, USA}
\altaffiltext{3}{Department of Physics and Astronomy, The Johns Hopkins University, 3701 San Martin Drive, Baltimore, MD 21218, USA}

\begin{abstract}
The Cold Spot on the Cosmic Microwave Background could arise due to a
supervoid at low redshift through the integrated Sachs-Wolfe effect.
We imaged the region with MegaCam on the Canada-France-Hawai'i
Telescope and present galaxy counts in photometric redshift bins.  We
rule out the existence of a 100Mpc radius spherical supervoid with
underdensity $\delta=-0.3$ at $0.5<z<0.9$ at high significance.  The
data are consistent with an underdensity at low redshift, but the
fluctuations are within the range of cosmic variance and the low
density areas are not contiguous on the sky.  Thus, we find no strong
evidence for a supervoid.  We cannot resolve voids smaller than 50Mpc
radius; however, these can only make a minor contribution to the CMB
temperature decrement.
\end{abstract}


\section{Introduction}
The nature of the Cold Spot on the Cosmic Microwave Background (CMB)
has been the source of broad speculation.  The 10\degr~diameter region
identified in \emph{Wilkinson Microwave Anisotropy Probe} (WMAP)
temperature maps \citep{WMAP1} is curious due to its pronounced
temperature and morphology. Extensive study of the region was first
motivated by the Spot's non-Gaussian properties
\citep{Vielva04,Cruz05}.  Under the standard cosmological model, the
primordial fluctuations on the CMB are homogeneous, isotropic and
described by a Gaussian random field.  In this scenario, the existence
of the Spot is unlikely at the $\sim$0.5\% level \citep{Cruz06},
although see \citet{Zhang09} for a critical view.  The mean
temperature decrement within a 5\degr~radius is $\Delta T=-100$\muK,
and it shows no systematic variation with frequency in WMAP data,
making it inconsistent with contamination from synchrotron or dust
emission \citep{Cruz06,rudnick07}.

Many sources of secondary anisotropies have been proposed to explain
the Cold Spot, including the integrated Sachs-Wolfe (ISW) and
Sunyaev-Zeldovich (SZ) effect, as well as more exotic physics
including textures arising in the early Universe.  Evidently there
is no sufficiently massive cluster in the local universe to produce an
SZ temperature decrement over such a large angular scale; however, the
other hypotheses have not been directly addressed through
observations.  \citet{Cruz08} review these possibilities.

\begin{figure}[th]
    \begin{center}
      \includegraphics{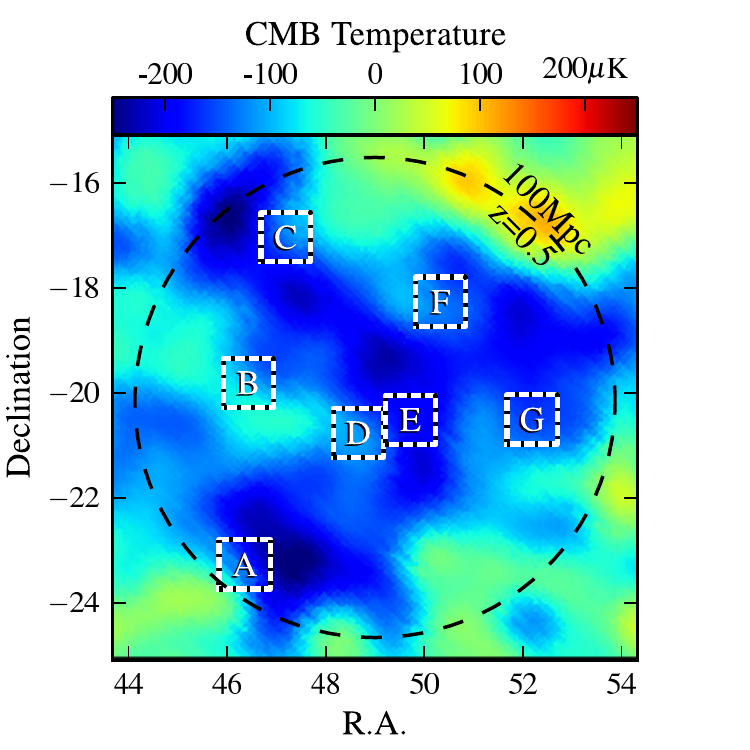}
    \end{center}  
    \caption{The 7 Cold Spot fields overlaying the WMAP linear
      combination (ILC) temperature map.  The dashed circle indicates
      the angular scale of 100Mpc radius at z=0.5.
    \label{fig:fields}}
\end{figure}

In this work, we address the question of whether a supervoid exists
along the line-of-sight.  A significant CMB decrement can be induced
by a time-varying gravitational potential through the integrated
Sachs-Wolfe effect (ISW) \citep{sachswolfe}.  On linear scales, CMB
photons traversing a large-scale void lose energy as the potential
decays under the accelerated cosmological expansion.  Non-linear
structure growth can also contribute to a cold imprint on the CMB
through the Rees-Sciama effect \citep{RS68}.

This scenario garnered interest due to the recent direct measurement
of voids imprinted on the CMB \citep{supervoids}.  The 10\muK~signal
was measured on 4\degr~scales and has been speculated to arise from
100Mpc scale underdensities.  Could a similar structure be responsible
for the Cold Spot?  Furthermore, \citet{rudnick07} found a
coincidental depression in source counts in the NRAO VLA Sky Survey
(NVSS), although the significance of this alignment has been debated
\citep{Smith08}.  Through an independent methodology, \citet{McEwen07}
found that the Cold Spot region contributes strongly to the positive
NVSS-WMAP cross correlation.  These findings motivate further
investigation of the void hypothesis.

Survey data covering the Cold Spot are limited.  The ISW signal arising
from the local Universe has been studied by \citet{maturi07} who
consider structures within 100Mpc.  They find no significant signal at
the location of the Cold Spot.  \citet{Francis09a} investigate the ISW
signal traced by 2MASS at $z<0.3$.  They detect an underdensity at the
Cold Spot, but find that it contributes to the temperature decrement
by only 5\%.

The Cold Spot is most prominent in a compensated filter.
\citet{Cruz06} use a Spherical Mexican Hat Wavelet with a scale radius
of 4.17\degr~and find a temperature decrement of $-16.09$\muK~in this
filter with a 1$\sigma$ range due to cosmic variance of 3.55\muK.
Assuming that a void contributes only within the positive range of the
wavelet, we need a $\sim$10\muK~contribution to reproduce the Spot on
top of a $1-2\sigma$ primary fluctuation on the CMB.

A 10\muK~temperature decrement requires a $\sim$100Mpc scale
underdensity.  A spherical underdensity with $\delta=-0.3$ must have a
radius of 200Mpc to produce the effect \citep{inouesilk07,Sakai08}.
For the purposes of this work, we compute the expected ISW signal
using the order-of-magnitude derivation presented by \citet{rudnick07}
which is in agreement with this result at low redshift.

To test the existence of a 100Mpc supervoid at $z<1$, we carried out
an imaging survey of the Cold Spot region with MegaCam on the
Canada-France-Hawai'i Telescope (CFHT).  Using galaxy counts, we
estimate the large-scale density distribution in photometric redshift
bins to $z=0.9$ and we compare our measurement to the expected
distribution from mean counts across the sky.  We assume a standard
WMAP5 \LCDM~cosmology with $\Omega_m=0.31$, $H_0=0.72~{\rm
  km~s^{-1}~Mpc^{-1}}$ and $\sigma_8=0.80$.

\begin{table*}[ht]
\begin{center}
\caption{Survey fields\label{table:fields}}
\begin{tabular}{|c|c|c|c|cccc|}
\tableline
  &       &           & Area      &\multicolumn{4}{|c|}{95\% Completeness (AB mag)}   \\
ID& R.A.  &  Dec.     & (sqr deg) & $g$& $r$& $i$& $z$     \\
\tableline
A & 03:05:26.5 & -23d15m21s &0.80 &22.7 & 21.6 & 22.4 & 20.9 \\
B & 03:05:45.3 & -19d48m07s &0.82 &22.8 & 22.8 & 22.5 & 21.0 \\
C & 03:08:45.8 & -17d01m35s &0.81 &23.2 & 23.4 & 22.4 & 20.8 \\
D & 03:14:40.1 & -20d45m16s &0.81 &22.6 & 23.2 & 22.4 & 20.5 \\
E & 03:18:53.2 & -20d30m25s &0.77 &22.4 & 23.4 & 22.4 & 21.1 \\
F & 03:21:20.0 & -18d15m23s &0.88 &23.9 & 23.3 & 22.1 & 21.5 \\
G & 03:28:45.3 & -20d29m45s &0.79 &23.1 & 23.0 & 22.1 & 21.1 \\
\tableline
\end{tabular}
\end{center}
\end{table*}

\section{Data}
\subsection{Observations}
We imaged the Cold Spot region with MegaCam on the
Canada-France-Hawaii Telescope (CFHT) in the optical filters $griz$.
The observations were taken by staff observers from Oct.-Nov. 2008.
The survey includes seven 0.8 deg$^2$ fields within 5\degr~of the Cold
Spot listed in Table \ref{table:fields}, and plotted on the sky in
Fig. \ref{fig:fields}.  The positions on the sky were chosen to avoid
bright stars.  Field F covers the NVSS depression in number counts
identified by \citet{rudnick07}.

\begin{figure}
    \begin{center}
      \includegraphics{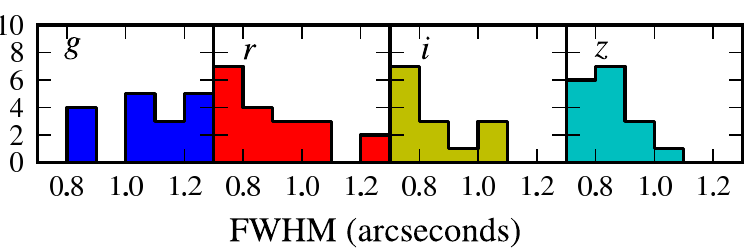}
      \includegraphics{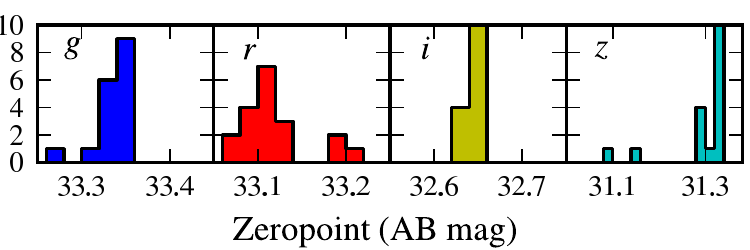}
    \end{center}  
    \caption{The data were taken under a variety of conditions.  These
      plots show the distribution in stellar FWHM (top) and
      photometric zeropoint (bottom) of the MegaCam images.
    \label{fig:fwhm}}
\end{figure}

The survey was designed to detect red galaxies to facilitate
photometric redshift estimation.  The nominal integration times were
1200, 1400, 1000 and 840 seconds in $g$,$r$,$i$,$z$, respectively.
Each integration was divided into two offset exposures.  More than two
exposures were obtained of some fields with sufficient quality to be
combined in the analysis.  Histogram plots of the image PSF FWHM and
zeropoints are presented in Fig. \ref{fig:fwhm}.

\subsection{Data reduction}
The MegaCam data were processed by the Elixer pipeline at CFHT. This
facility applies standard bias subtraction, flat fielding and fringing
corrections to the images.  Elixer also determines a photometric
solution based on standard fields observed over the same nights.

We used the Astromatic software \scamp~to determine an astrometric and
relative photometric solution for each image and \swarp~to bring the
images into pixel alignment \citep{scamp}.  Cosmic ray hits were detected and masked
using the LACosmic IDL code~\citep{lacosmic}.  Lastly, \swarp~was used to
generate median stacked images.

Due to varying observing conditions, the data have a range in stellar
point spread functions (PSFs) from 0.7-1.2\arcsec~full-width at
half-maximum (FWHM).  For robust aperture color measurements, we
generated PSF-matched image stacks.  We convolved each image with
FWHM$<1\arcsec$ with a PSF-matched kernel.  The kernel was determined
by first fitting a Moffatt PSF model to high signal-to-noise stars for
each of the 36 chips within the MegaCam focal plane.  The kernel was
then found that would result in a standard profile with FWHM=1\arcsec.
In practice, the resulting convolved image has a FWHM$>1\arcsec$ due
to tails in the PSF, and we find that PSF FWHM of the convolved images
fall in the range from 1-1.2\arcsec.  The masked regions, including
defective pixels and cosmic ray hits, were also enlarged to account
for the convolution.

We generated catalogs using SExtractor software \citep{tractor}.  The
convolved images were processed in two-image mode to match the
photometric apertures in each band.  Source detection was performed on
the $i$ images.  We use the SExtractor {\scshape mag\_auto} magnitudes
to measure galaxy magnitudes and colors.

We investigated the completeness limits of the catalogs by adding
artificial galaxies to the raw images and repeating the processing
steps.  The galaxies were modeled with $R^{1/4}$ exponential profiles
with effective radii of $R_e=(0.5'',1.0'',1.5'')$ matching the range
in surface brightness of real sources.  An artificial galaxy was said
to be detected if the measured SExtractor magnitude was within 1
magnitude of the true value.  The magnitude completeness limits for
each field are listed for effective radius $R_e=1.0''$ in Table
\ref{table:fields}.  Of note is field A which is shallow in $r$.  The
completeness limits are only directly relevant in $i$ band since we do
not require significant detections in the other filters.

Field masks were constructed to exclude gaps in the detector, bad
columns and the halos and diffraction spikes surrounding bright stars.
Mask processing was facilitated by Mangle2 \citep{mangle}.

\subsection{Archival data}
To determine the expected galaxy counts in the Cold Spot fields, we
examined archival data available from the CFHT Legacy Survey (CFHTLS).
We obtained catalogs and masks from data release 4, which includes 16
fields from the wide and deep surveys with $griz$ photometry.  The
field locations and areas are listed in Table \ref{table:cfhtls}.
Field D1 overlaps W1 making the total survey area 9.8 sqr deg.

Both the CFHTLS Wide and Deep survey images are deeper than those of
the Cold Spot.  To account for this in our analysis, we degrade the
photometry with Gaussian noise to match our Cold Spot fields.

\begin{table}
\begin{center}
\caption{CFHTLS Fields\label{table:cfhtls}}
\begin{tabular}{lcc}
\tableline
   &        &  Area     \\
ID & RA Dec & (sqr deg) \\
\tableline
W1 &02h20m --04d12m &  6.69 \\
W2 &09h05m --02d23m &  0.70 \\
D1 &02h25m --04d29m &  0.77 \\
D2 &10h00m +02d12m  &  0.79 \\
D3 &14h19m +52d24m  &  0.83 \\
D4 &22h15m --17d43m &  0.80 \\
\tableline
\end{tabular}
\end{center}
\end{table}

We can directly compare our own data with archival MegaCam fields with
the caveat that the \emph{i}-band filter was replaced since the
beginning of the Legacy project.  Our data were obtained with the new
filter, designated i.MP9702, while the archival data are from the
original i.MP9701 filter.  The filter transmissions are sufficiently
different that a color transform of order 0.1 magnitudes must be
applied to compare the two data-sets.

When comparing \emph{i}-band galaxy counts, we convert to the new
i.MP9702 filter using the relation
$i_{9702}=i_{9701}+0.078(r-i_{9701})$ derived from synthetic red
galaxy photometry.  We generate synthetic stellar and galaxy
photometry using transmission functions obtained from
S. Gwyn\footnote{The filter transmission functions are available
online at
\url{http://www4.cadc-ccda.hia-iha.nrc-cnrc.gc.ca/megapipe/docs/filters.html}}.

\subsection{Sample selection}
\begin{figure}
    \begin{center}
      \includegraphics{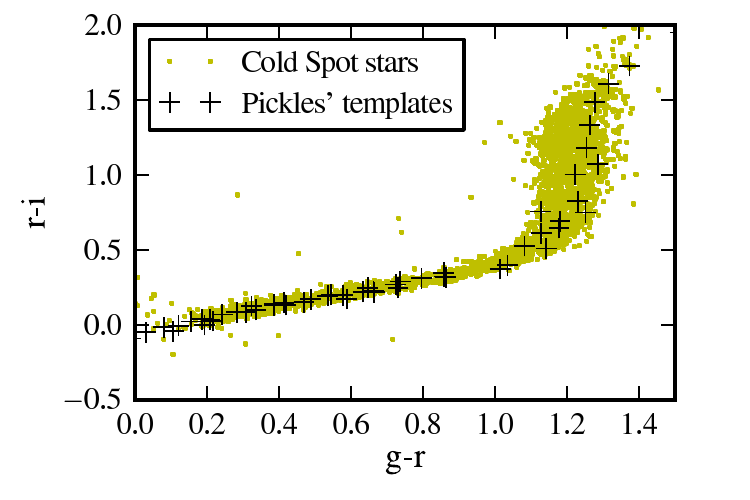}
    \end{center}  
    \caption{We apply color corrections to align the stellar loci in
      $griz$ color space, illustrated here by the $g-r$,$r-i$ plot.
      Plotted are 2900 stars from the 7 Cold Spot fields.  The plus
      symbols mark synthetic photometry from Pickles' stellar SEDs.
    \label{fig:stars}}
\end{figure}

We found that due to residual zeropoint variations, additional
corrections were warranted to obtain internally consistent stellar
colors.  Within each field, we fit the stellar loci in \emph{griz}
color space, see Fig. \ref{fig:stars}. Offsets of $\sim$0.05
magnitudes were applied to align the loci to model fits using the
Pickles' stellar library \citep{Pickles}.  The typical errors in the
alignment fits are: $\sigma_{g-r}=0.02$, $\sigma_{r-i}=0.01$ and
$\sigma_{i-z}=0.01$ magnitudes.  We applied color corrections to both
Legacy and Cold Spot fields.  We apply a dust extinction correction to
extragalactic sources from \citet{Schlegel}.

We classified sources as stars or galaxies based on the half-light radius
in $i$ measured by SExtractor.  We used the criteria presented in
\citep{Coupon} in which the cutoff is determined from a Gaussian fit to the
distribution of radii for sources in the field.  We classified sources
to a limiting magnitude of $i_{9702}=21.5$; all fainter sources were
treated as galaxies.

\begin{figure}
    \begin{center}
      \includegraphics{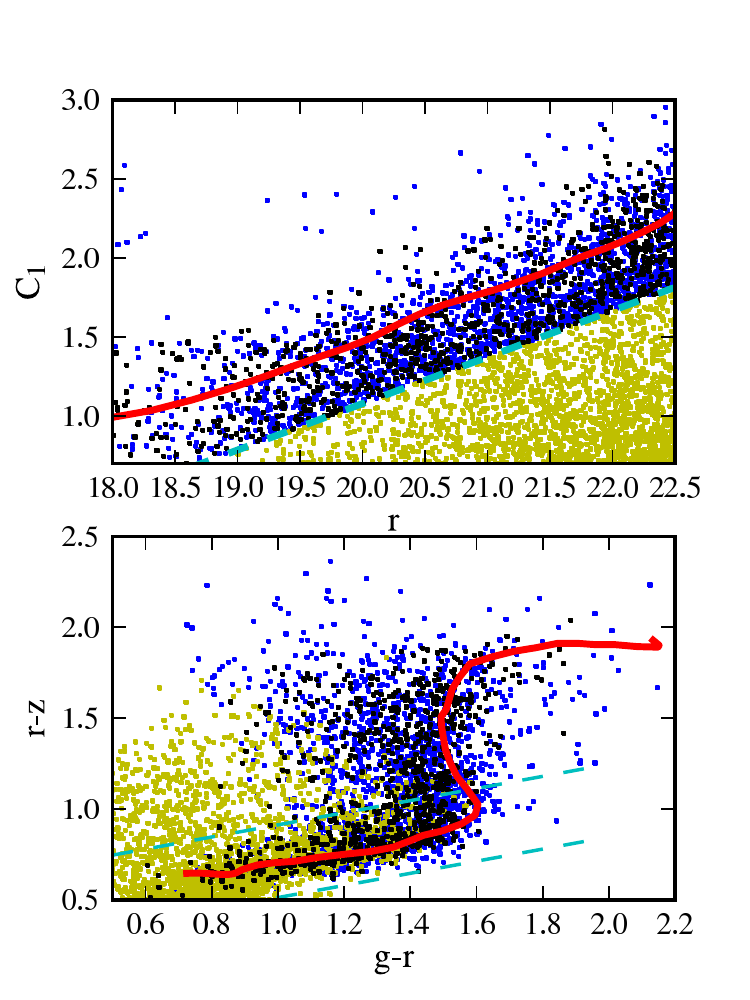}
    \end{center}  
    \caption{The red galaxy sample was selected in $grz$ color space.
      At top: a luminosity cut was applied in $r$ following the
      evolutionary track.  Bottom: cuts are made in $grz$ space.
      Overplotted is an evolutionary track for a passively evolving
      galaxy.
    \label{fig:colorsel}}
\end{figure}

We constructed a sample of red galaxies in $grz$ space.  Due to the
required photometric transform, we do not use the \emph{i}-band for
galaxy selection, ensuring that no selection bias arises between the
Legacy and Cold Spot fields.  The selection criteria are based on a
passively evolving galaxy model \citep{bc03} and are illustrated in
Fig. \ref{fig:colorsel}; the cuts are as follows.  
(1.) We defined a rotated color space $c_1,c_2$ aligned with the red
galaxy track at $z<0.4$ and impose a luminosity cut following the red
sequence: $17<r<22.5$, $c_1=0.7(g-r)+3(1-0.7)(r-z-0.38)$, $c_2=(r-z) -
(g-r)/3 - 0.38$, $c_1>0.294r-4.8$.  (2.) The following cuts include
low redshift ($z<0.4$) galaxies: $|c_2|<0.2$, $c_1<2.0$.  (3.) High
redshift galaxies are included with: $c_2>0.2$, $0.5<g-r<2.5$,
$r-z<2.5$.  The conservative $r<22.5$ cut was used to ensure that
the sources were well detected in $r$, $i$ and $z$.

\section{Magnitude-number relation}
\begin{figure}
    \begin{center}
      \includegraphics{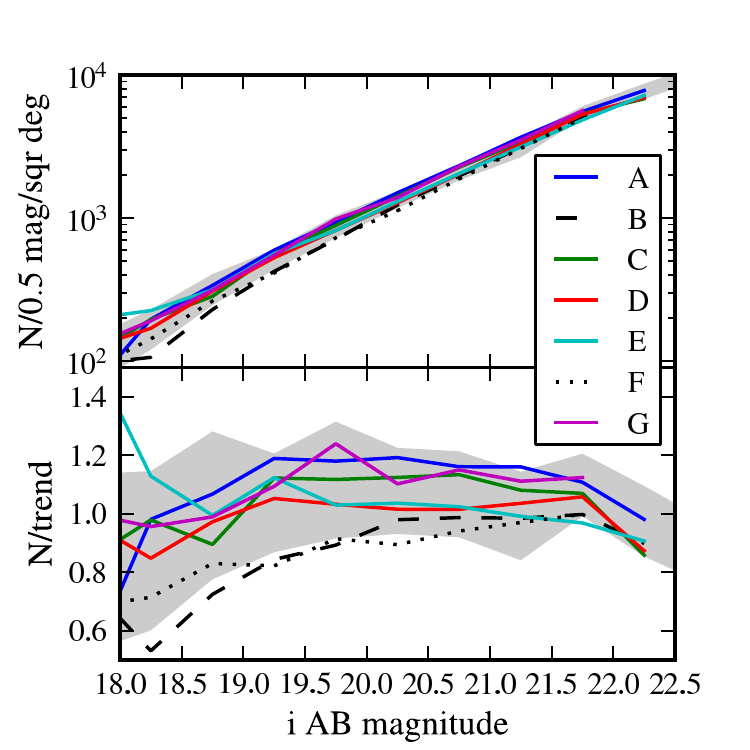}
    \end{center}  
    \caption{The differential galaxy counts as a function of $i$-band
      magnitude.  The top panel shows the counts per square degree in
      0.5mag bins.  The lines represent the 7 Cold Spot fields, A-G.
      The shaded region was measured from 15 CFHTLS fields and
      represents the range of cosmic variance.  The bottom panel plots
      the same counts divided by the exponential trend.  All fields
      are consistent with the CFHTLS data.
    \label{fig:icounts}}
\end{figure}

A coarse view of the galaxy distribution can be obtained from the
differential galaxy count function with magnitude.
Fig. \ref{fig:icounts} shows the \emph{i}-band counts we measure in
the Cold Spot fields including all sources classified as galaxies; the
shaded region was measured from 16 CFHTLS fields over 9.8 sqr deg and
represents the range of cosmic variance at the $\sim2-\sigma$ level.

The Cold Spot fields are consistent with the Legacy fields.  Of note
are fields B and F, which show low number counts.  Interestingly, F
covers the low density spot in NVSS.  The fields were not imaged under
poor conditions, so there is no evidence that this is due to a
systematic effect in completeness.  Fields B and F are separated on
the sky, so it is unlikely that the two measurements are related
through a single large underdensity.

\section{Redshift distribution}
We use photometric redshifts to examine the galaxy distribution along
the line of sight to the Cold Spot.  The redshifts of early type
galaxies can be estimated from broadband photometry based on the
position of the 4000\AA~break \citep{Connolly95}.  The $gri$ colors
constrain the redshift at $0.2<{\rm z}<0.4$ while the $riz$ colors
primarily constrain the redshift at higher z, $0.4<{\rm z}<1.0$.

However, biases can arise from color-redshift degeneracies limiting
the accuracy of photometric redshifts and introducing artificial
features in the derived redshift distributions.  We mitigate this by
carrying out identical analysis procedures on both the Cold Spot and
CFHTLS catalogs which we construct to have uniform photometric
properties.  We use the CFHTLS fields to find the expected mean
counts, including the survey selection function and redshift error.
We then use the redshift error estimates to deconvolve the photo-z
distribution and find the underlying galaxy counts.  The homogeneity
of the data makes the procedure resilient to photo-z biases that may
arise, for instance, from photometric uncertainty or from the limited
spectroscopic calibration set.  The methodology and results are
described in the subsequent sections.

\subsection{Photometric redshifts}
We estimated galaxy redshifts using the \emph{griz} photometry.  We
used the template fit photo-z algorithm \eazy~v1.00 \citep{eazy}.  The
\eazy~code computes fits to photometric colors using linear
combinations of galaxy SED templates and allows for magnitude-redshift
priors.  We used the built-in \eazy~template set which consists of six
principal SED components with the appropriate MegaCam filter
transmission curves.  We added a broad $r$-band magnitude prior
calibrated from the training set data as well, which is described
below.

The spectroscopic sample consists of 928 galaxies from multiple
surveys that overlap the CFHTLS fields, see Table \ref{table:zspec}.
The redshift ranges and numbers listed include only sources within our
red galaxy sample.  We divided the sample into two sets of roughly 465
galaxies each to form separate training and validation sets.

Due to the luminosity cut there is a strong trend of galaxy magnitude
with redshift within the red galaxy sample. To take advantage of this,
we used the training sample to construct a magnitude-redshift prior.
We adopted a functional form for the prior of $p(z|m)\propto
z^{a}\exp[-(z/z_0)^a]$ with parameters $a(m)$ and $z_0(m)$ tuned
to match the mean and variance of the data.  The mean and variance
were fit as linear functions of magnitude; however, we artificially
broadened the prior by scaling the variance by a factor of 4.  

The model zeropoints can be inaccurate due to uncertainties in the
instrumental transmission and photometric system.  To improve the
quality of the photo-z's, we varied the color zeropoint offsets to
optimize the fits.  We iteratively ran the \eazy~code on the training
set while varying the color zeropoint offsets using a simulated
annealing optimization algorithm.  The quality of the fit was measured
by the root-mean-square error with 5-$\sigma$ outliers removed.

\begin{figure}
    \begin{center}
      \includegraphics{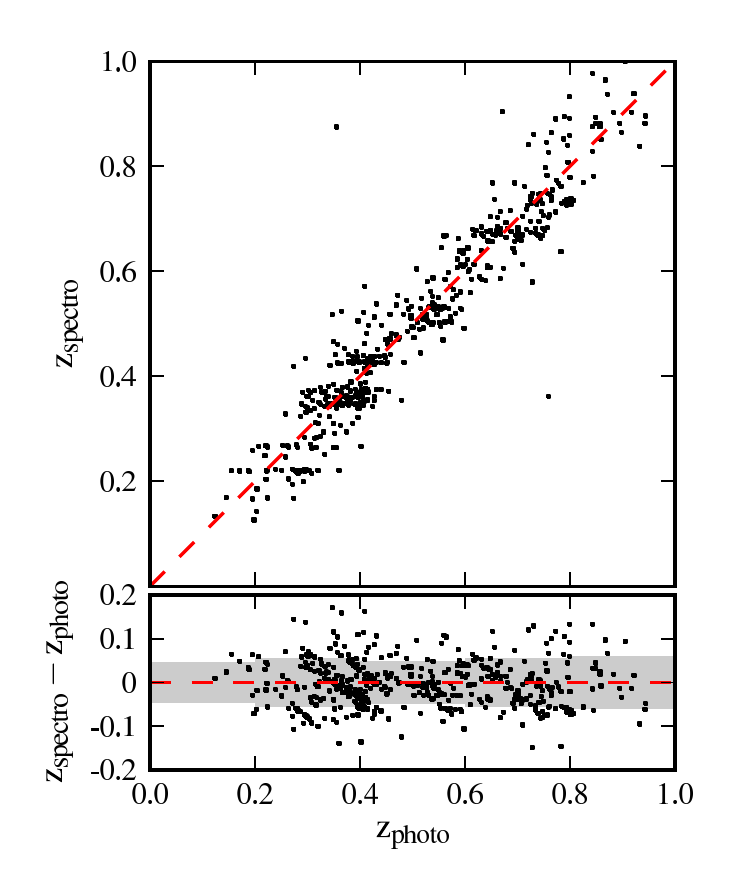}
    \end{center}  
    \caption{The photometric redshift precision demonstrated on the
     validation set of $\sim465$ galaxies.  The photometry was
     degraded to model the photometric errors in the Cold Spot data.
     The shaded regions show $1-\sigma_{RMS}$ ranges.
\label{fig:photoz}}
\end{figure}

The results on the validation set is shown in Fig. \ref{fig:photoz}.
Here, Gaussian noise has been added to the CFHTLS photometry to model
the Cold Spot data.

We used the full training set including artificial noise to estimate
the photo-z error distribution.  The error varies with redshift and
may be biased.  We fit the distribution in redshift bins of 0.1 with a
two-Gaussian model. The error kernels are illustrated in
Fig. \ref{fig:sf}.

\begin{table}
\begin{center}
\caption{Photo-z Training Set\label{table:zspec}}
\begin{tabular}{lclr}
\tableline
Survey & CFHTLS Field & z-Range (Median z)& N \\
\tableline
ZCOSMOS\tablenotemark{a}  & D2 & 0.12--1.00 (0.47) & 678  \\
DEEP2\tablenotemark{b}    & D3 & 0.20--0.98 (0.55) & 125  \\
VVDS\tablenotemark{c}     & D1 & 0.14--0.88 (0.59) & 111   \\
SDSS LRG\tablenotemark{d} & D2,D3 & 0.31--0.38 (0.35) & 14  \\
\tableline
\tablenotetext{1}{\citet{zcosmos}}
\tablenotetext{2}{\citet{deep2a}}
\tablenotetext{3}{\citet{vvds}}
\tablenotetext{4}{\citet{sdss}}
\end{tabular}
\end{center}
\end{table}

\subsection{Monte Carlo inversion}
We modeled the underlying galaxy distribution with a Monte Carlo
Markov Chain method.  The model redshift distribution was constructed
with four redshift bins from z=0.1-0.9, with $\Delta z=0.2$.  We chose
this coarse binning to minimize systematic errors, including
uncertainties in the photo-z error kernel.  The number of galaxies in
each bin was treated as a parameter in the MCMC, giving four degrees
of freedom.  We evolved the chain using a random walk according to the
Metropolis-Hastings algorithm.

At each step in the iteration, the true number of galaxies in each
redshift bin was specified.  The initial condition was set to the
result of a Richardson-Lucy deconvolution and subsequent steps were
then chosen in a random walk manner from a Gaussian proposal
distribution.  Additionally, proposed steps were required to give a
positive number of galaxies in each bin.

The Monte Carlo photo-z distribution was constructed by distributing
the model galaxies according to the photo-z error kernel.  The
likelihood of this distribution was then computed according to the
Poisson distribution, $f(N_i;\lambda_i)$, with the mean, $\lambda$,
set by the expected photo-z number counts from the archival data.
Each proposed step was either accepted or rejected according to a
likelihood ratio where the probability of accepting the $k^{th}$ step
depends on the ratio of likelihoods, as,
\begin{equation}
p = \min\{\prod_i f(N_{i,k};\lambda_i)/f(N_{i,k+1};\lambda_i),1\}
\end{equation}

The chain converges quickly because only neighboring bins are strongly
correlated.  We carried out 50000 steps for each field and dropped the
first 1000 iterations to ensure that the results are not affected by
the starting condition.

\begin{figure}
    \begin{center}
      \includegraphics{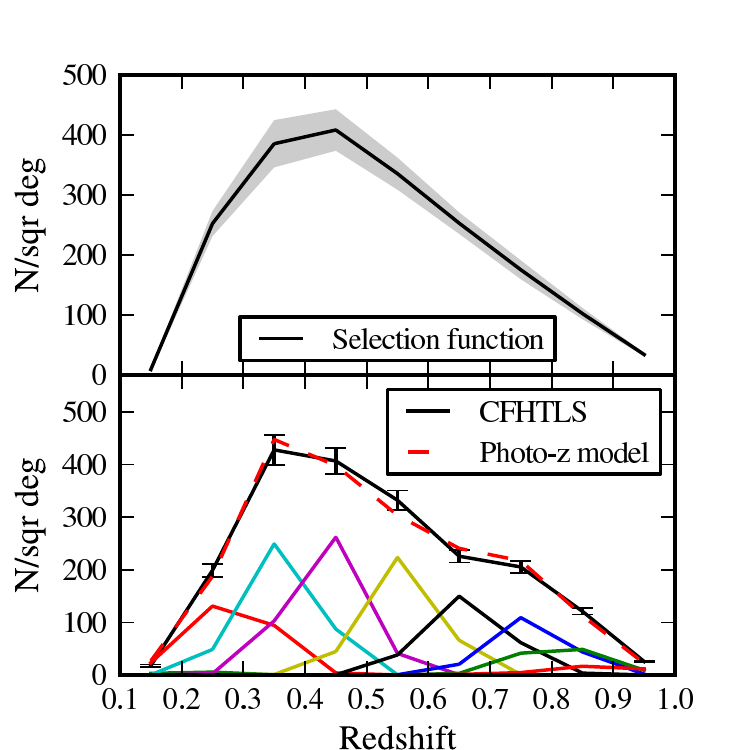}
    \end{center}  
    \caption{Top: a spline fit of the deconvolved selection function.
      Bottom: the photo-z distribution (solid line) is well fit by the
      model selection function convolved with the photo-z error
      kernels (dashed line).  The distribution decomposed into the
      photo-z error kernels is plotted along the bottom.  The error
      bars represent uncertainty in the fit arising from cosmic
      variance.
    \label{fig:sf}}
\end{figure}

We used the same MCMC procedure on the archival fields to determine
the selection function.  In this case, the galaxy counts from all 15
archival fields were summed and normalized by the total area.
Fig. \ref{fig:sf} illustrates the photo-z redshift distribution, along
with a spline fit of the deconvolved distribution.

\begin{figure*}
    \begin{center}
     \mbox{(a)\includegraphics{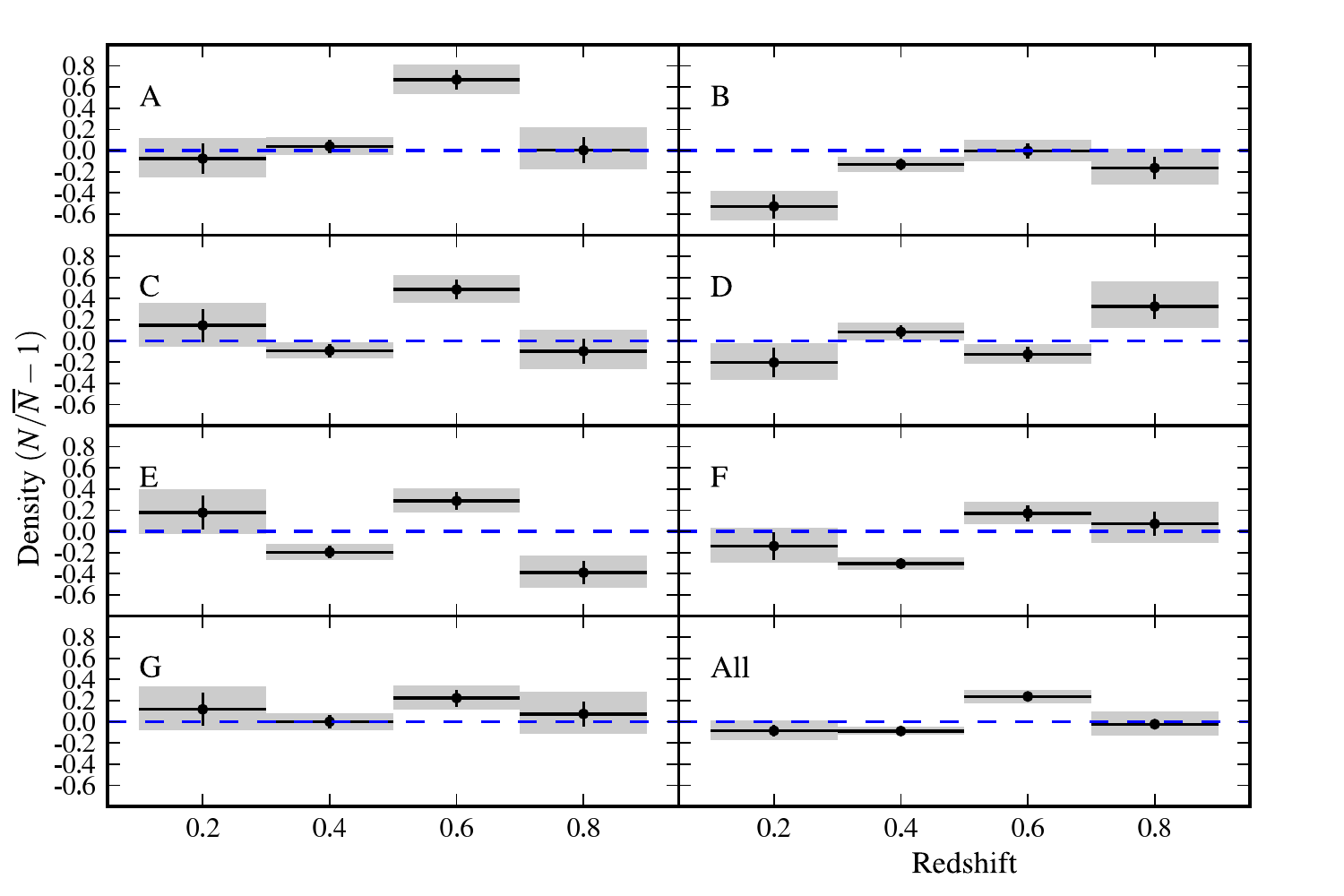}}
     \mbox{(b)\includegraphics{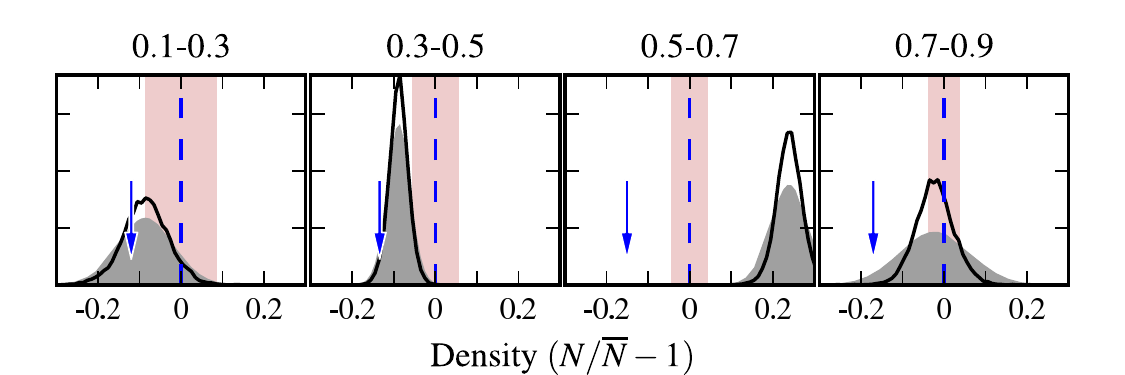}}
     \mbox{(c)\includegraphics{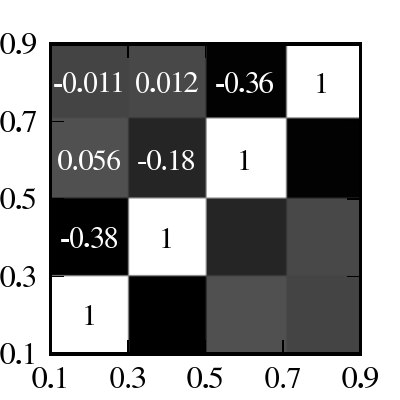}}

    \end{center}  
    \caption{Plotted are the corrected redshift number density
      distributions based on Monte Carlo modeling.  Panel (a) shows
      the result from each of the 7 fields (\emph{A-G}), as well as
      the mean distribution derived by summing the galaxy counts
      (\emph{All}).  The error bars give 68\% marginalized likelihood
      ranges.  Panel (b) shows the marginalized likelihood
      distribution of the density for the four redshift bins in the
      combined measurement.  The filled histograms include systematic
      uncertainties in the selection function.  The vertical shaded
      region is the 1-$\sigma$ range of cosmic variance and the arrow
      marks a 200Mpc diameter supervoid with underdensity
      $\delta=-0.3$.  The typical normalized covariance between
      redshift bins is illustrated in Panel (c).  Neighboring bins are
      anti-correlated by 20-40\%.
    \label{fig:zdist}
}
\end{figure*}

We normalized the galaxy counts by the selection function, and express
the result as the overdensity, $N/\overline{N}-1$.  The redshift
distributions for the seven individual Cold Spot fields are shown in
Fig. \ref{fig:zdist}, along with the combined measurement from summing
all of the fields.  The likelihood function was found from the MCMC
runs and arises from Poisson error.  On top of this, there is a
systematic uncertainty in the selection function determination.  The
plots also include a systematic error in the $r$ magnitude zeropoint
of 0.05.  This affects the two high redshift bins where the selection
function is steeply falling.  The added error is 1.5\% in bin 3 and
6.1\% in bin 4; in comparison, the Poisson errors in these bins are
3.7\% and 4.7\%, respectively.  In the next section, we consider how
color shifts affect the photo-z distribution as well.

\begin{figure}
    \begin{center}
      \includegraphics{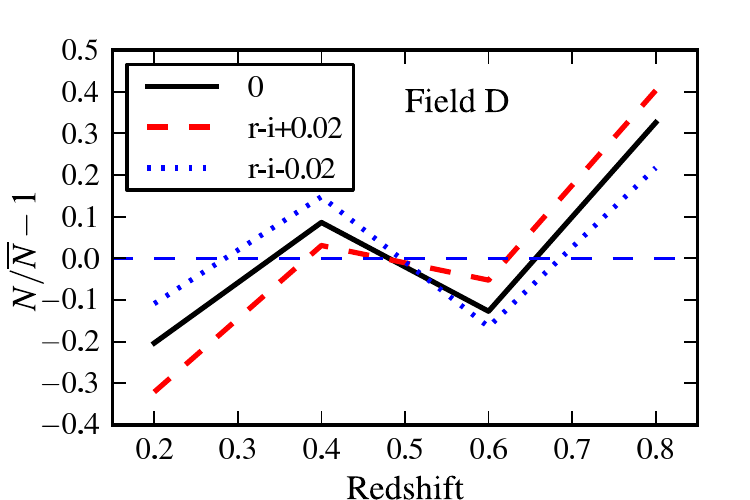}
    \end{center}  
    \caption{To check the effect of systematic color offsets on the
      redshift distribution, we recomputed the photo-z's with
      zeropoint offsets in $r$ of $\pm.02$ magnitudes.  The plot shows
      the resulting measurement in field D.  The distribution is
      shifted towards higher (+0.02) or lower (-0.02) redshift by the
      perturbations.
    \label{fig:robust}}
\end{figure}

\begin{figure}
    \begin{center}
      \includegraphics{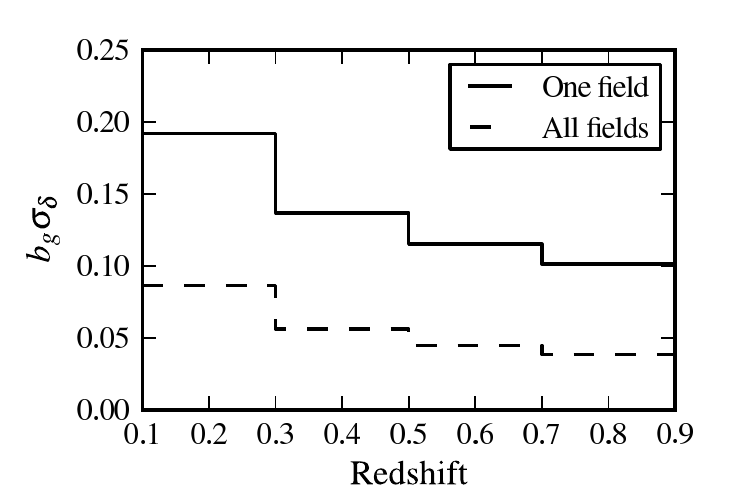}
    \end{center}  
    \caption{The expected $1\sigma$ range in Gaussian density
      fluctuations in each redshift bin for one field and for all
      fields combined.  We assume a linear galaxy bias of 1.5.
    \label{fig:cosmic}}
\end{figure}

\begin{figure*}
    \begin{center}
      \includegraphics[scale=1]{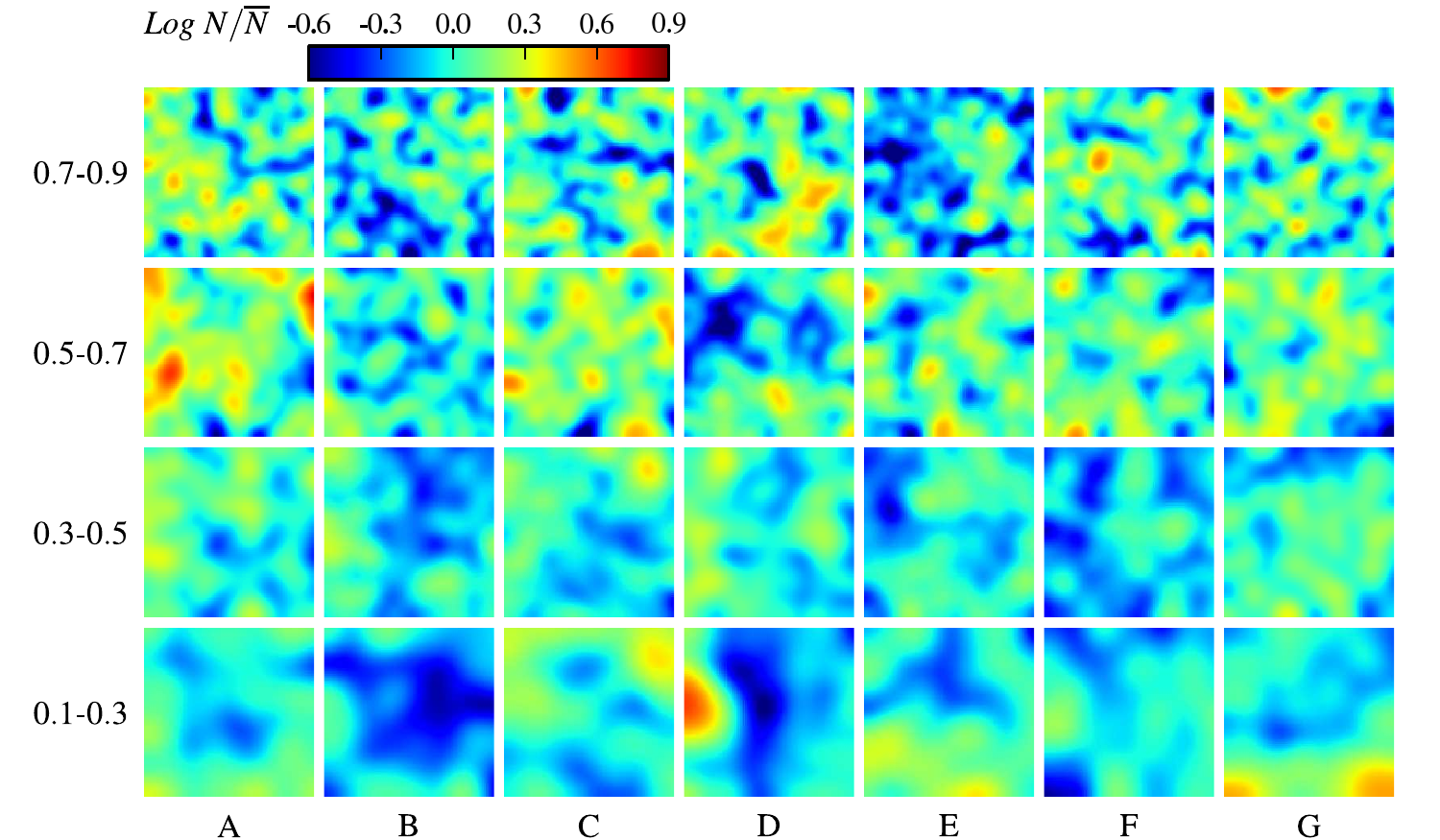}
    \end{center}  
    \caption{ Large structures are evident in the projected densities
      including two clusters at z=0.6 in field A, and an apparent void
      in field D in the same redshift bin.  The density in each
      redshift bin is smoothed with a Gaussian kernel with a FWHM of
      2Mpc.  Each plot is 1\degr~square.  North is up and RA increases
      to the left.  Masked regions excluding chip gaps and bright
      stars have been filled with a uniform distribution of galaxies
      according to the selection function.
    \label{fig:slices}}
\end{figure*}

\subsection{Systematic uncertainties}
A systematic variation in color or magnitude of sources between fields
can significantly affect the derived photometric redshift
distributions.  Calibration errors in individual fields will tend to
average out in the combined analysis, but a systematic difference
between the archival and Cold Spot fields could strongly bias our
conclusions.  There are two primary issues: the magnitude zeropoint
and color offsets.

The magnitude zeropoint affects the number of galaxies at high
redshift through the $r$ magnitude limits and luminosity cut.  We
expect the photometric calibration from CFHT Elixer system to be
better than 0.05 magnitudes and the offset due to the extinction
correction is of the same order.  We find that a shift of 0.05
magnitudes changes the sample size by 5\% in the high redshift bins.
This is is smaller than the Poisson error in number counts in a single
field, but could bias the selection function.  Based on the
magnitude-number counts, as well as the redshift counts, we find no
evidence for significant discrepancies.  The zeropoint also affects
the photometric redshifts through the magnitude prior, but this
dependence is minor.

Variations in color between fields is more serious because these can
redistribute galaxies in the photo-z distribution.  The stellar loci
alignment can be made to a precision of $r-i=.01$ magnitudes.  The
uncertainty in $g-r$ is greater, but has less of an effect on the
photo-z estimates.  Systematic uncertainties may come from the
Galactic extinction correction as well, which has a similar order of
$g-r\sim.02$.  To test the effects of color shifts, we added an $r-i$
color offset of $\pm0.02$ magnitudes and recomputed the redshift
distribution.  The result is a shift of the photo-z distribution to
higher or lower redshift affecting number counts in bins by 10\%, see
Fig. \ref{fig:robust}.  Though we do not expect such a large shift,
this does limit the constraints we can put on the density especially
in the lowest and highest redshift bins.

The redshift distributions in a number of fields show linear trends in
Fig. \ref{fig:zdist}, particularly fields C, D and E.  In field D, due
primarily to the first and last bins, the distribution shows a rising
trend with increasing redshift, while field E shows the opposite
trend.  These features could be due to residual color zeropoint
errors.  There are also apparent oscillations in the number counts in
fields D and E.  The anti-correlations between bins may contribute to
this, but these are properly accounted for in the MCMC procedure and
represented in the marginalized error bars.  A prominent feature is a
significant overdensity at z=0.6 that appears in four of the seven
fields and dominates the combined signal.  This feature could be an
artifact of the selection function perhaps due to an inaccurate
photo-z model.  We investigated how these features depend on the $r$
magnitude limit.  With a deeper $r$ limit of 23.0, the redshift
distributions are not significantly altered and the oscillations
persist in fields D and E; however, the overdensity at z=0.6 is
reduced, perhaps due to broader redshift errors.  With a brighter
limit of $r=22.0$, we detect few galaxies in the high redshift bin,
but the prominent features in the distribution remain including the
z=0.6 overdensity.  We conclude that there are hints of systematic
trends in the redshift distributions, but no significant features that
can be addressed with the calibration methods available.  We find that
many of the features do indeed correspond to real structures,
validating the redshift distributions.

\section{Discussion}
Structures are evident in the individual Cold Spot fields.  The
projected densities in bins of width $\Delta z=0.2$ are plotted in
Fig. \ref{fig:slices}.  There is an excess of sources at $z=0.6$ in
field A, and we have confirmed by eye that there are indeed two
clusters in the field.

Field B has a very significant underdensity of 60\% in the first bin.
This underdensity is confirmed in $i$ magnitude counts
(Fig. \ref{fig:icounts}) which are low for $i_{AB}<20$.  Field F, which
overlaps the NVSS depression identified by \citet{rudnick07}, shows
low magnitude counts as well and is underdense in the two low redshift
bins by 10-30\%.  Thus, we conclude that the NVSS feature is due to an
underdensity at $z<0.5$.

In field D, the underdensity at z=0.6 is resolved in
R.A.-Dec. projection, suggesting that it is a real structure.  This
void is in the Northern part of field D, which has a width of
$\sim20$Mpc.  We ran the 3D Voronoi tessellation-based void-finder
\zobov\ \citep{zobov} on all Cold Spot and CFHTLS fields, and this
void has the largest density contrast of all voids found.  However, it
was necessary to severely distort the pencil beam into a compact shape
(a cube) to use a Voronoi tessellation on the galaxies effectively,
obscuring the physical meaning of this finding.  The void is not seen
in the adjacent field E, which has an overdensity in this bin, making
it unlikely that field D is part of a larger supervoid.

The expected range in density due to cosmic variance is plotted in
Fig. \ref{fig:cosmic} for both a single field and the combined
measurement.  We assume a linear galaxy bias $b_g=1.5$.  The
underdensity at z=0.8 in field E is $-0.39$; this is a 3.8$\sigma$
deviation in terms of cosmic variance.  The variations within fields
are larger than expected: many of the fields deviate by $>2\sigma$ in
at least one bin, and have a greater amplitude in the highest redshift
bins.  We find similar variations in the CFHTLS fields.  We attribute
the additional variance to systematic sources particularly affecting
the lowest and highest bins.  These issues make a robust void
detection difficult in any single bin.

Additionally, void detections are limited by cosmic variance.  Due to
the limited sky coverage, a typical deviation due to small structures
within a single field could be large enough to be consistent with a
supervoid especially in the low redshift bins.  However, we can detect
a supervoid by looking for coherent structures across many fields.
For instance, a 100Mpc radius void would reduce the number counts in
most of the fields.  Such a void with $\delta=-0.3$ would affect the
galaxy density by $b_g\delta\sim-.45$.  In a bin at z=0.6 with width
$\Delta z=0.2$, the reduction in counts would be 15\%.  The arrows in
Fig. \ref{fig:zdist}(b) mark the expected underdensity due to this
void in each redshift bin.

Using the combined likelihood distributions we can test the case of a
supervoid with radius 100Mpc and $\delta=-0.3$.  The density
measurements in the four redshift bins and the expected decrement for
this supervoid model are listed in table \ref{table:limits}.  We
compute a likelihood, $p(<\delta)$, of measuring a density less than
that of the supervoid based on the realizations in the MCMC chain.

At z=0.2, we measure an underdensity of 8.5\% which is consistent with
the supervoid.  The underdensity at z=0.4 is not deep enough to match
the void model at the 97\% confidence level, neglecting systematic
uncertainties in the selection function.  We find an overdensity at
z=0.6, and the measured number density in the last bin, at z=0.8, is
inconsistent with the supervoid at the 99.7\% level.

We note that the 140Mpc empty void model proposed by \citet{rudnick07}
is inconsistent with our measured redshift distribution at the level
of 1 in 50000 in each redshift bin and can be ruled out.

\begin{table}
\begin{center}
\caption{Supervoid model limits\label{table:limits}}
\begin{tabular}{|c|c|c|l|}
\tableline
Redshift & {\tiny$N/\overline{N}-1$ } & {\tiny$(N/\overline{N}-1)_{model}$} & $p(<\delta)$\\
\tableline
0.1-0.3 & -0.085  & -0.12 & 0.25   \\
0.3-0.5 & -0.088  & -0.13 & 0.029  \\
0.5-0.7 &  0.24  & -0.15 & $<1e-5$    \\
0.7-0.9 & -0.022  & -0.17 & 0.0026 \\
\tableline
\end{tabular}
\end{center}
\end{table}

Do the underdensities measured at $z<0.5$ provide evidence for a
supervoid?  The deviation of 8.5\% in the first bin is only a
$1.0\sigma$ fluctuation in terms of cosmic variance.  This alone
suggests that it is an unlikely source of the Cold Spot since it is a
typical fluctuation.  It is predominantly detected in only fields B, D
and F, but not in adjacent field E.  Furthermore, the only fields that
also show a low magnitude-number relation are B and F.  It is likely
that a large underdensity would not be homogeneous, but the data do
not provide overwhelming evidence for a coherent large structure.  The
underdensity at z=0.4 is interesting because it is measured in four of
the seven fields.  It is a $1.6\sigma$ deviation with respect to
cosmic variance.  This provides a hint of a large structure, but more
sky area is required to make a robust measurement.

Smaller voids may extend over a subset of the fields.  At z=0.2,
fields B, D and F are particularly underdense with an angular extent
of 5\degr~or 50Mpc at this redshift.  However, the linear ISW
temperature decrement predicted for a 50Mpc diameter volume is
0.3\muK.  At z=0.4, a structure with the same angular extent would be
100Mpc in diameter and produce a 0.9\muK~ decrement.  It is unlikely
that even having many of these modest supervoids along the line of
sight could produce the Cold Spot feature.

The significant overdensity at z=0.6 measured in four of the fields
could represent a massive structure.  The 25\% overdensity is
consistent with a 200Mpc diameter overdensity with $\delta=0.5$.  To
linear order, this would induce a 5\muK~hot spot on the CMB.  This
suggests that the primary anisotropy on the CMB is even colder than
observed.  Overdensities can induce cold spots due to the non-linear
Rees-Sciama effect which comes to dominate at high redshift
\citep{Sakai08}, but this is thought to be important only at $z>1$.

\section{Conclusions}
We investigated the distribution of galaxies at $z<0.9$ in the
direction of the CMB Cold Spot.  We detect an underdensity at low
redshift, $0.1<z<0.3$, that is consistent with the density expected
from a supervoid.  This underdensity appears to be present in 2MASS as
well \citep{Francis09a}.  However, due to the limited sky coverage of
our survey, we cannot draw a definite conclusion regarding the
existance of a coherent supervoid structure.  Our measurements
disfavor a supervoid with $\delta=-0.3$ at z$=0.3-0.5$, and we can
rule out a supervoid at $z=0.5-0.9$.  We must be cautious in
interpreting the density measurements due to possible systematic
shifts in the selection function, especially in the near and far
redshift bins.

In \citet{supervoids,Granett09} we found the imprint of supervoids in
SDSS to be stronger than predicted from linear ISW. This suggests that
a more modest supervoid with $\delta\gtrsim -0.3$ could be the origin of the
Cold Spot decrement.

Significant progress will be made in understanding the Cold Spot with
wide-area large-scale structure surveys including Pan-STARRS-1
\citep{panstarrs}.  Additionally, polarization data from the Planck
mission \citep{Planck} may provide additional information on the
intrinsic CMB fluctuations in this area.

\acknowledgments 
We thank Roy Gal for sharing telescope time and Adrian Pope for
contributing his expertise.  Some of the results were derived with
CosmoPy (\url{http://www.ifa.hawaii.edu/cosmopy}), Healpix
\citep{healpix}, Healpy and the Astromatic software suite
(\url{http://astromatic.iap.fr}). We acknowledge the use of the LAMBDA
archive (\url{http://lambda.gsfc.nasa.gov}).  We are grateful for
support from NASA grant NNG06GE71G, NSF grant AMS04-0434413, and the
Pol\'anyi Program of the Hungarian National Office for Research and
Technology (NKTH).  This work is based on observations obtained with
MegaPrime/MegaCam, a joint project of CFHT and CEA/DAPNIA, at the
Canada-France-Hawaii Telescope (CFHT) which is operated by the
National Research Council (NRC) of Canada, the Institute National des
Sciences de l'Univers of the Centre National de la Recherche
Scientifique of France, and the University of Hawaii.  We wish to
recognize and acknowledge the very significant cultural role and
reverence that the summit of Mauna Kea has always had within the
indigenous Hawaiian community.  We are fortunate to have the
opportunity to construct observatories on this mountain.

\bibliographystyle{hapj}
\bibliography{refs}

\end{document}